\documentclass[twocolumn,showpacs,superscriptaddress,aps,pre]{revtex4-1}

\usepackage{graphicx}
\usepackage{amsmath}
\usepackage{dcolumn}
\usepackage{bm}
\usepackage{soul}
\usepackage[normalem]{ulem}
\usepackage{color}
\usepackage[caption=false]{subfig}
\usepackage[T1]{fontenc}
\usepackage{float}
\usepackage[12h=true]{scrtime}
\usepackage[dvipsnames]{xcolor}
\usepackage{mathtools, nccmath}
\usepackage{textcomp}
\newcommand{\rin}{r_\text{in}}
\newcommand{\rout}{r_\text{out}}
\newcommand{\B}[1]{{\bm{#1}}}
\newcommand{\Michael}[1]{\noindent \color{red}} 

\begin{document}
	\title{Disorder Induced Nonlinear Mode Coupling and Symmetry Breaking in Amorphous Solids}
\author{Avanish Kumar}
\affiliation{Dept. of Chemical Physics, The Weizmann Institute of Science, Rehovot 76100, Israel}
\author{Itamar Procaccia} 
\affiliation{Dept. of Chemical Physics, The Weizmann Institute of Science, Rehovot 76100, Israel}
\affiliation{Center for OPTical IMagery Analysis and Learning, Northwestern Polytechnical University, Xi'an, 710072 China}
\author{Murari Singh}
\affiliation{Dept. of Physics, Indian Institute of Technology, Tirupati 517619, India}

%%%%%%%%%%%%%%%%%%%%%%%%%%%%%%%%%%%%%%%%%%%%%
\begin{abstract}
Applying {\em very small} purely radial strains on amorphous solids in radial geometry one observes elastic responses that break the radial symmetry. Without any plasticity involved, the responses indicate nonlinear mode coupling contributions even for minute strains. We show that these symmetry-breaking responses are due to disorder, typical to amorphous configurations. The symmetry breaking responses are quantitatively explained using the classical Michell solutions which are excited by mode coupling. 
\end{abstract}
%%%%%%%%%%%%%%%%%%%%%%%%%%%%

\maketitle
%%%%%%%%%%%%%%%%%%%%%%%%%%%%  

\section{Introduction}

\begin{figure}
		\includegraphics[width=0.35\textwidth,angle=-90] {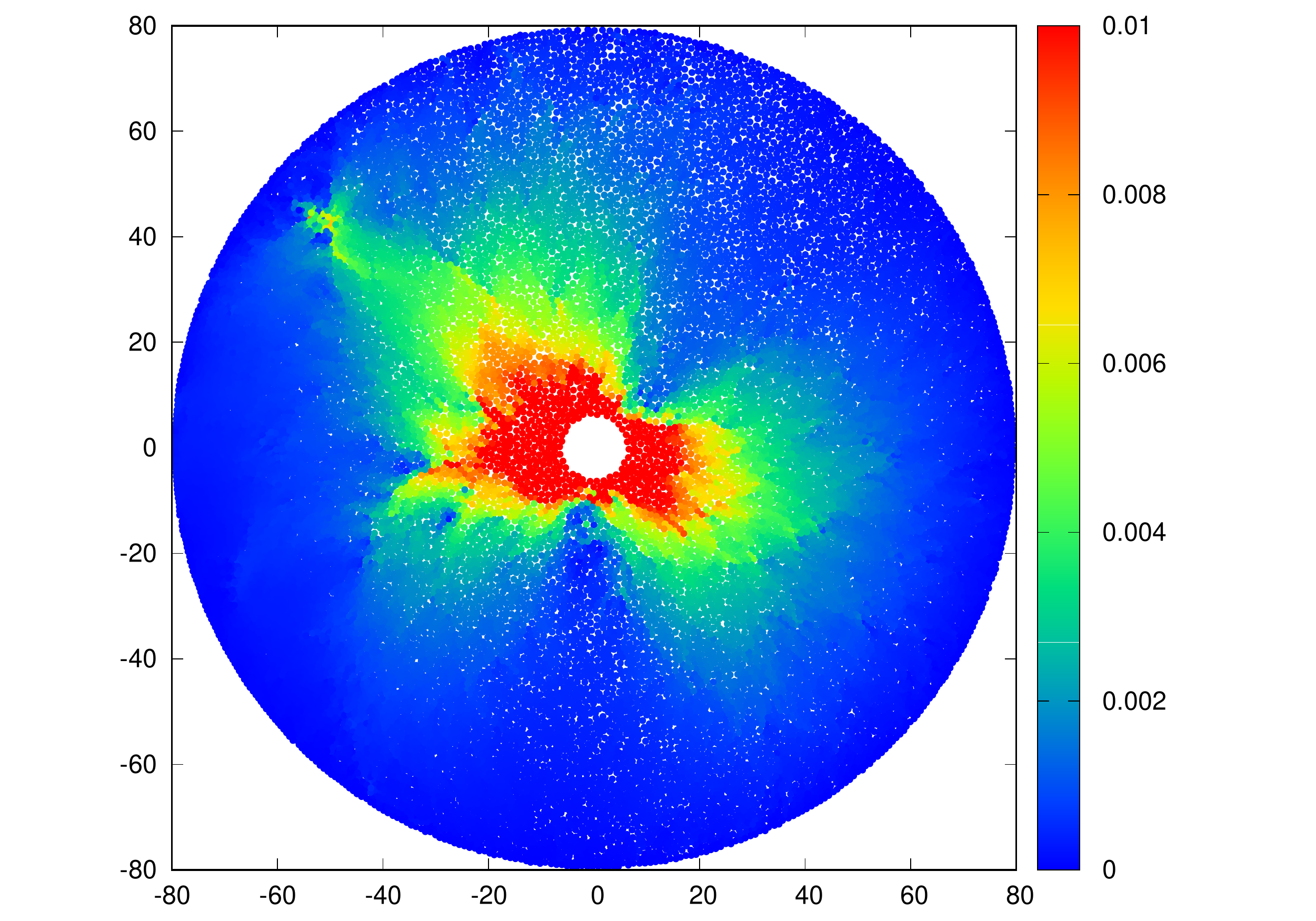} 
		\caption{Magnitude of the displacement field resulting form a minute
		purely radial inflation $r_{\rm in} \to r_{\rm in}+\delta$ where $r_{\rm in}=5$,
			$r_{\rm out}=80$ and $\delta=10^{-6}$. The concern of this paper is the non-radial response seen in this image.  }
		\label{simulation}
		\end{figure}

It is very customary in physics to assert that the effects of small perturbations on a given system can be
faithfully analyzed using linear or linearized theories. This is certainly correct in
mechanics, where small external strains on any given solid are expected to induce stresses and displacement fields that are perfectly predictable by linear elasticity theory \cite{Landau,02Bar}. But this may not be the case when the solid in question is amorphous. Even with very small strains, the spatial disorder that characterizes amorphous solids is not ``small" in any sense, and can result in responses that usually appear in classical solids only at much larger magnitudes of strain. In this paper we demonstrate this using a classical model of amorphous solids consisting of point particles interacting via Lennard-Jones forces with a distribution of interaction diameters.  $N$ particles are confined to an annulus with inner radius $r_{\rm in}$
and outer radius $r_{\rm out}$. The inner radius can be inflated, and the inflation is strictly radial, $r_{\rm in} \to r_{\rm in}+\delta$. In this paper we choose minute values of $\delta/r_{\rm in}$, of the order of $10^{-7}$. The outer boundary is rigid, leading to vanishing radial component of the displacement field. Having such a small inflation in mind one expects to see only purely radial displacement field throughout the sample, since only nonlinear effects can lead to the breaking of the radial symmetry. The interesting and perhaps surprising typical result that simulations discover are shown in Fig.~\ref{simulation}. 
%%%%%%%%%%%%%%%%%%
%%%%%%%%%%%%%%%%%%%%%%%%%%%%%%%%%%%%%%
The actual displacement field is not radial. We should stress that every configuration
of the amorphous solid exhibits different symmetry-breaking, depending on the realization of the random
structure of the solid, as explained below. The inescapable conclusion is that disorder induces mode coupling to non-radial modes which are exposed and analyzed in this Letter. 

The structure of this Letter as follows: in Sect.~\ref{simu} we describe the numerical experiments and extract the components of the displacement field that are responsible for the symmetry-breaking. To understand the nature of these components we turn in Sect.~\ref{Theory} to the Michell solutions of the radial elasticity problem \cite{02Bar,99Mic}, and demonstrate their relevance for the phenomenon under study. In Sect.~\ref{recon} we demonstrate that the symmetry-breaking modes in the displacement field can be faithfully reconstructed with a few low-order non-radial Michell solutions.  We offer a summary and a discussion in Sect.~\ref{discuss}. In this discussion we assign the mode coupling seen in the numerics to the existence of a disorder length, which typically increases when the pressure in the amorphous solid decreases. With the radius of the inner boundary $r_{\rm in}$ exceeding this length, the mode-coupling effects reduce to nothing, and the expected purely elastic response is regained. To complete the assignment of the symmetry breaking to the presence of disorder, we repeat the simulations using perfectly ordered configurations. In these cases no nonlinear mode-coupling is found. Finally we comment on the relevance of the present findings to the modeling of mechanical responses of amorphous solids using elasto-plastic models \cite{97SLHC,98HL,18NFMB}, arguing that disorder-induced nonlinear effects cannot be overlooked.

\section{Numerical Simulations}
\label{simu}
\subsection{system preparation}
\label{prep}
In the simulations we construct an annulus
with two rigid walls, with an inner radius $\rin$ and outer radius $\rout$. The annulus is then filled up with $N$ point particles put in random positions in the area $A=\pi (\rout^2-\rin^2)$. The number of particles is chosen such that the density of the equilibrated glass (as described below) has a value $\rho=N/A$. In this section
we use  a standard poly-dispersed model of $N$ particles of mass $m=1$ \cite{19BFFSS}. The binary interactions are
\begin{eqnarray}
	&&\phi(r_{ij}) = \epsilon\left(\frac{\sigma_{ij}}{r_{ij}}\right)^{12} +C_0 +C_2\left(\frac{r_{ij}}{\sigma_{ij}}\right)^2+C_4\left(\frac{r_{ij}}{\sigma_{ij}}\right)^4\nonumber \\&& \epsilon\!=1, C_0\!=\!-1.92415, C_2\!=\!2.11106, C_4\!=\!-0.591097 \ . \label{IPL}
\end{eqnarray}
This potential is cut-off at $r=1.5\sigma$ with two smooth derivatives. The unit of energy is $\epsilon$ and Boltzmann's constant is unity.
The interaction length was drawn from a probability distribution $P(\sigma)
\sim 1/\sigma^3$ in a range between $\sigma_{\rm min}$ and $\sigma_{\rm max}$ such that the mean $\bar \sigma=1$:
\begin{eqnarray}
	&&\sigma_{ij} =\frac{\sigma_i+\sigma_j}{2}\Big[1-0.2\Big|\sigma_i-\sigma_j\Big|\Big],\nonumber\\ 
	&&\sigma_{\rm max} =1.61\ , \sigma_{\rm min}=\sigma_{\rm max}/2.219 \ .
\end{eqnarray}
The units of mass and length are $m$ and $\bar \sigma$ (the average $\sigma$). The parameters are chosen to avoid crystallization. The system is thermalized at some ``mother temperature" $T_m$ using Swap Monte Carlo and then 
cooled down to $T=0$, using conjugate gradient methods. The interaction between the point particles and the two walls are of the same form Eq.~(\ref{IPL}), where $r_{ij}$ and $\sigma_{ij}$ are replaced by the distance to the wall and by $\sigma_i$. 

Once the system is mechanically equilibrated with the total force on each particle smaller than $10^{-8}$, we inflate the inner radius $\rin$, forcing a radial displacement of magnitude $d_0$ as 
reported below. After inflation the radius of the inner circles is $\rin+d_0$. It should be stressed that our inflations are instantaneous, not quasi-static. We made sure that no particle gets trapped in the inflated disk; all our particles are confined between the inner and outer boundaries. This of course limits the degree of inflation in our simulations. After inflation we mechanically equilibrate the system again by the conjugate gradients, and then measure the displacement of each particle \{$\B d_i\}_{i=1}^N$, comparing the two equilibrated configurations \{$r_i,\theta_i\}_{i=1}^N$  before and after inflation. 
%%%%%%%%%%%%%%%%%%%%%%%%%%%%%%%%%%%%%%%%%%%%%%%%%%%%%%%%%%%
\subsection{Numerical Results}
\begin{figure}
	\includegraphics[width=0.35\textwidth,angle=-90] {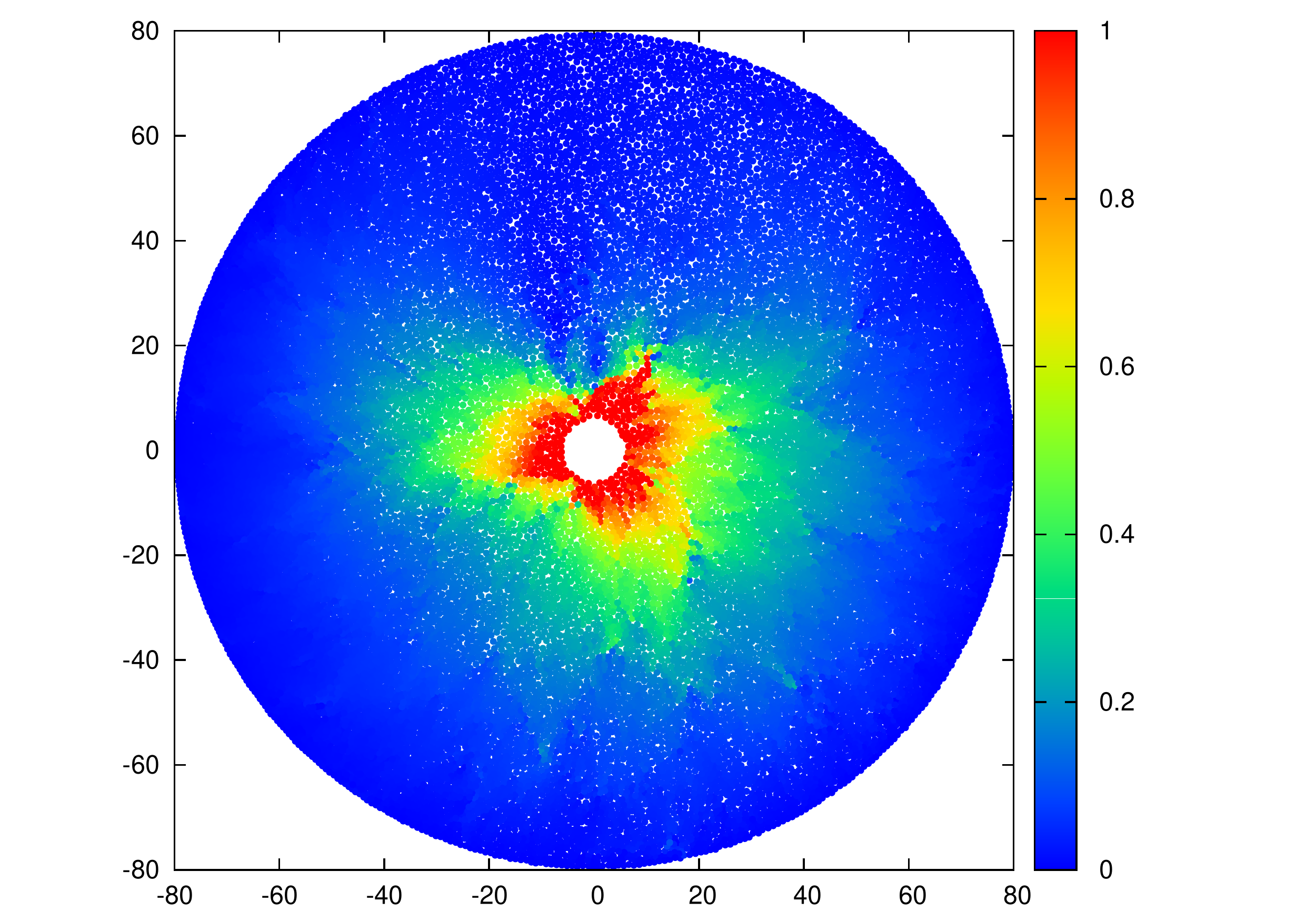}
		\includegraphics[width=0.35\textwidth,angle=-90] {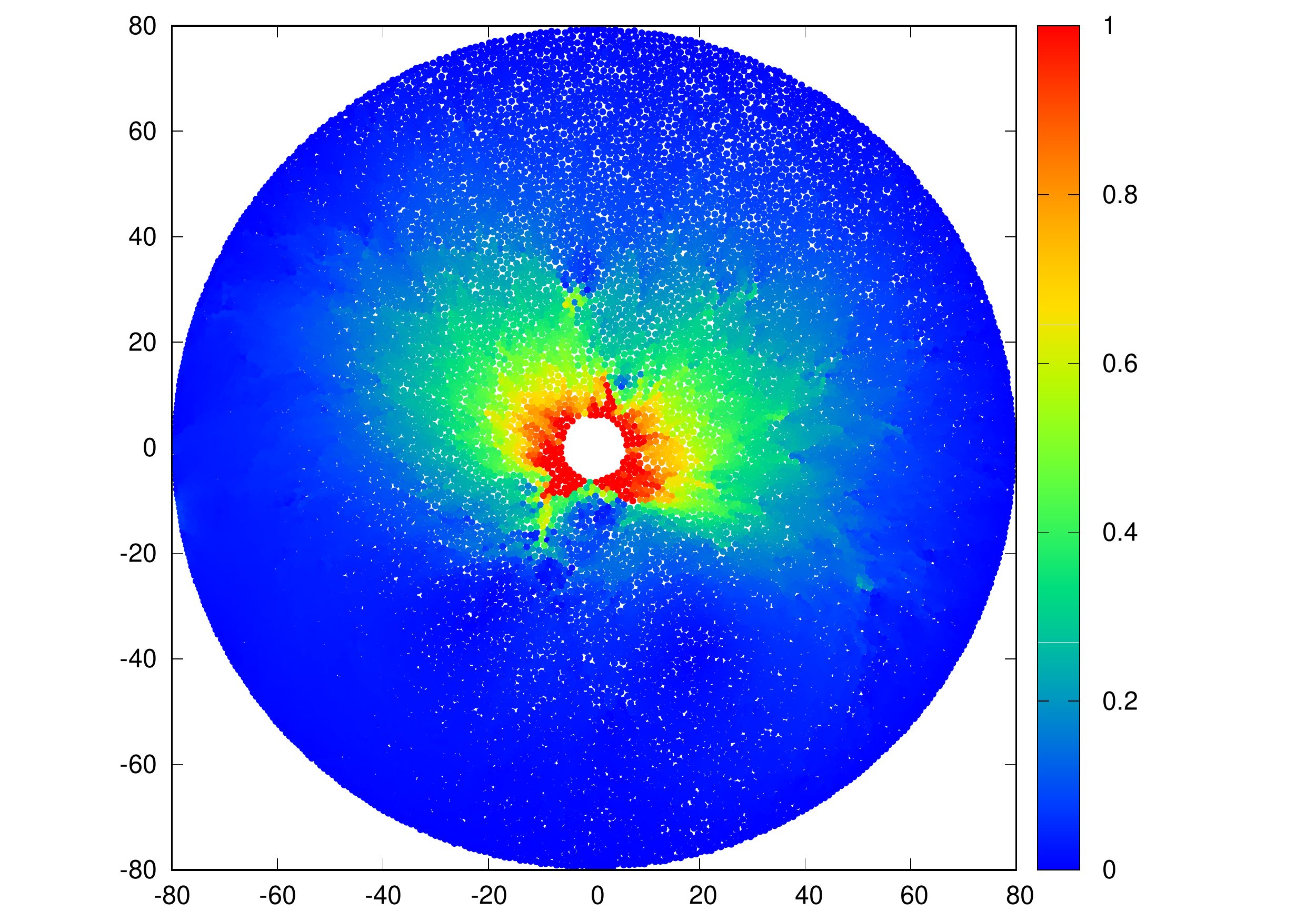}  
	\caption{Two further examples of the magnitude of the displacement field resulting form a minute
		purely radial inflation $r_{\rm in} \to r_{\rm in}+\delta$ where the parameters are all identical to those of Fig.\ref{simulation}. The point to notice is the variability of results for identical protocols, a reflection of the material disorder.}
	\label{sim2}
\end{figure}

Typical plots of the magnitude of the displacement field are shown, in addition to
Fig.~\ref{simulation}, in Fig.~\ref{sim2}.
%%%%%%%%%%%%%%%%%%
%%%%%%%%%%%%%%%%%%%%%%%%%%%%%%%%%%%%%%
The conditions and protocol leading to the results shown in Fig.~\ref{sim2} are identical to those in Fig.~\ref{simulation}. The difference is that each time the configuration was recreated starting from random initial conditions. Thus it is clear that the randomness in the resulting amorphous solid is responsible for the precise expression of the symmetry-breaking. Below we analyze and characterize
the non-radial components of the response using the classical Michell solutions. Section \ref{recon} demonstrates the reconstruction of the symmetry-breaking modes in the displacement field using non-radial Michell solutions. 

\subsection{Data Aanlysis}

\begin{figure}
	\includegraphics[width=0.35\textwidth] {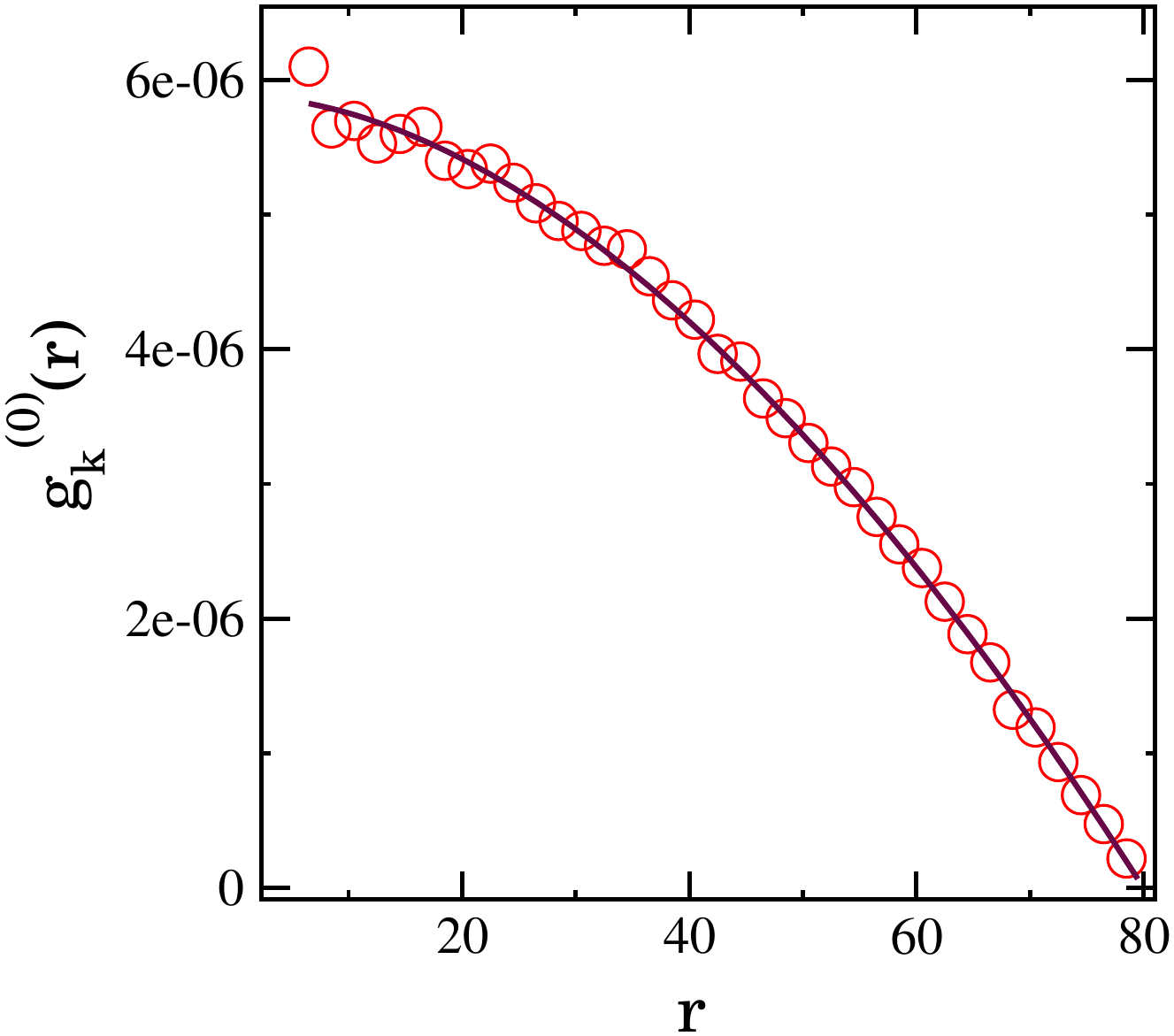}
	\caption{The numerically computed function $g_k^{(0)}(r)$ presented by the open dots. The continuous line is $I_0(r)$ defined by Eq.~(\ref{S10}). }
	\label{iso}
\end{figure}

\begin{figure}
		\includegraphics[width=0.35\textwidth] {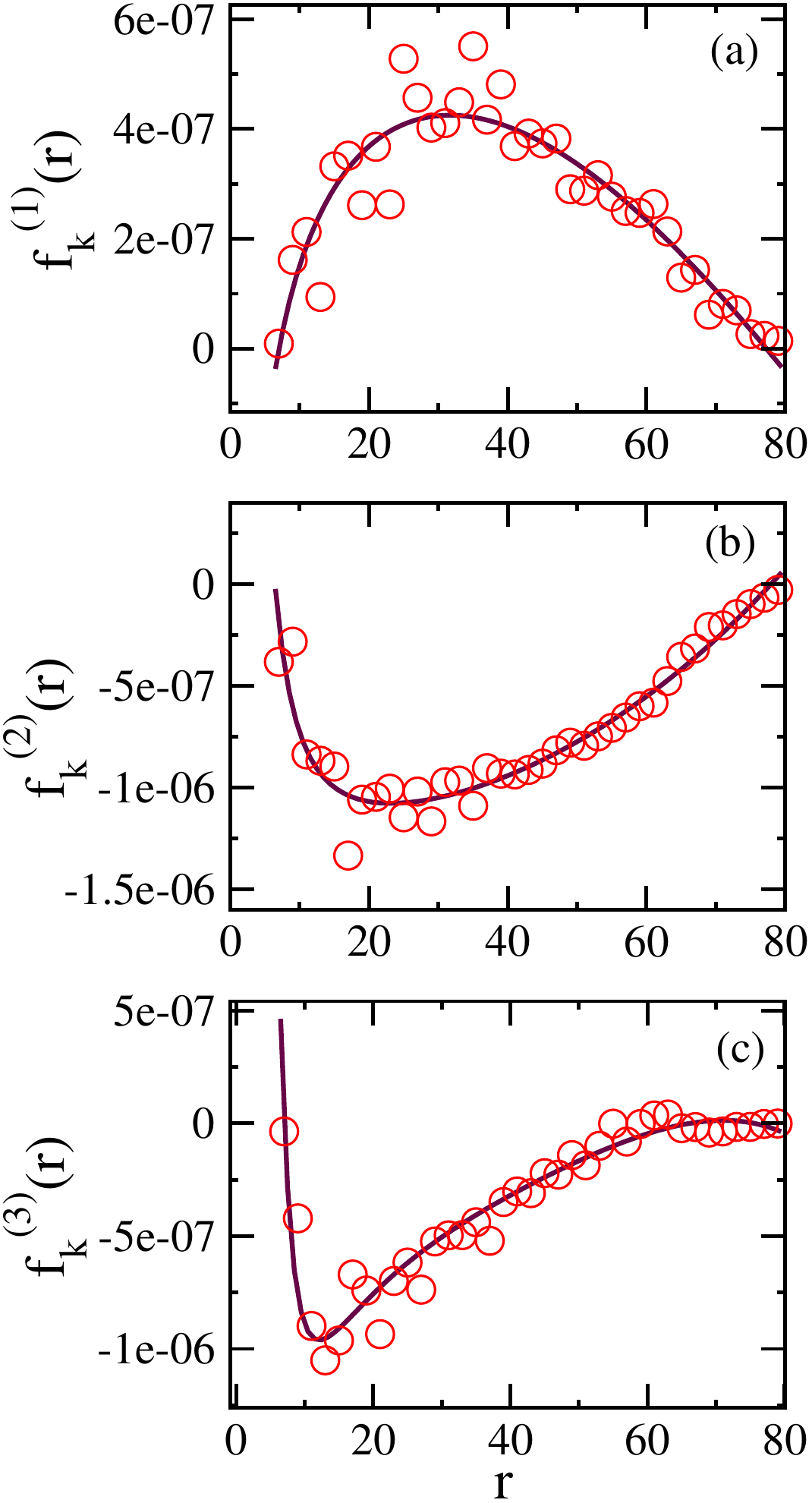}
		\caption{The function defined in Eqs.~(\ref{deffg}) for $n=1,2$ and 3. The continuous lines are $I_m$ of Eq.~(\ref{S11}) for $m=1,2$ and 3. }
		\label{sin}
	\end{figure}

\begin{figure}
			\includegraphics[width=0.35\textwidth] {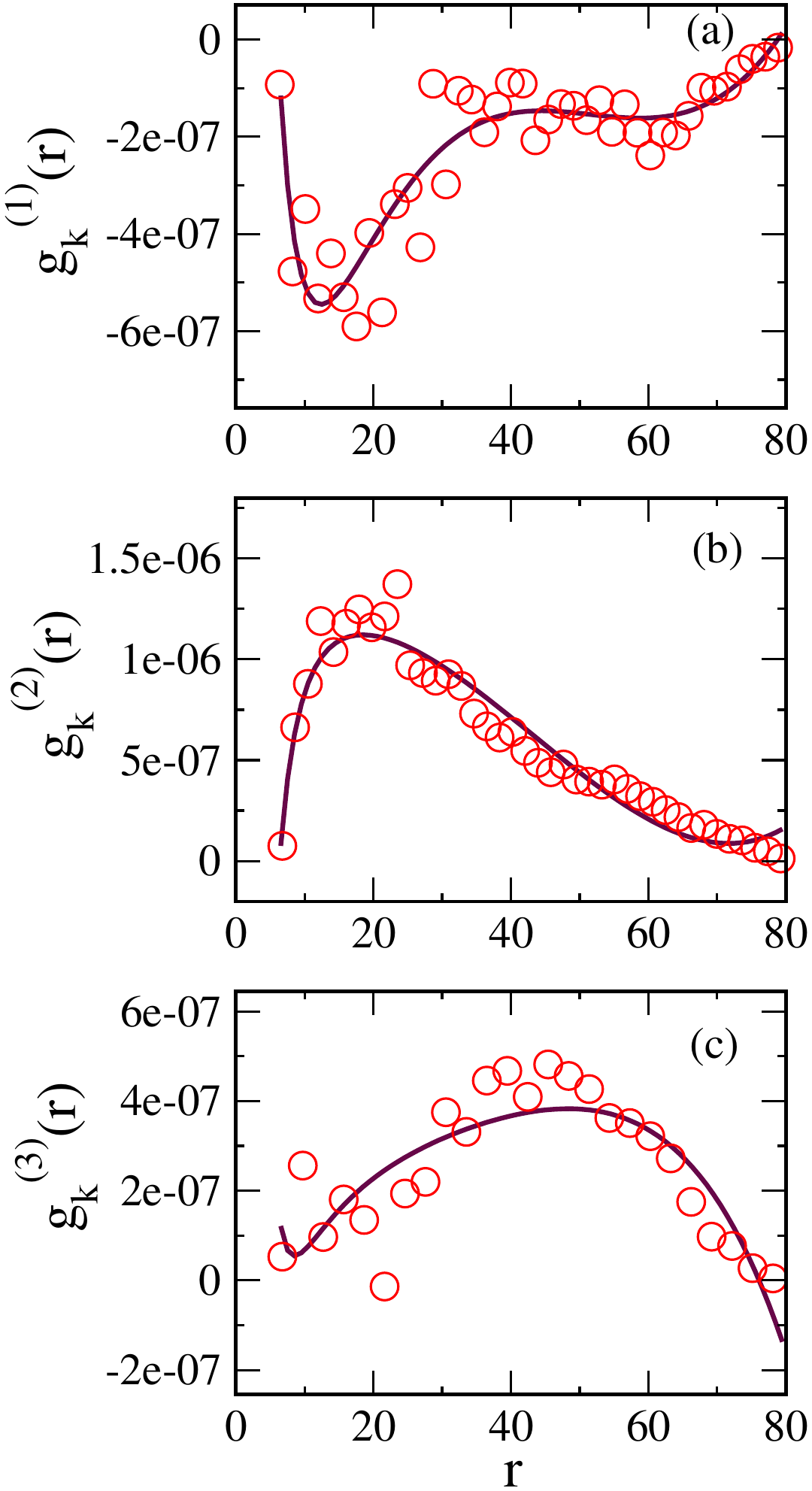}
			\caption{The function defined in Eqs.~(\ref{deffg}) for $n=1,2$ and 3. The continuous lines are $J_m$ which are the analogs of Eq.~(\ref{S11}) but for the cosine modes with  $m=1,2$ and 3.}
			\label{cos}
			\end{figure}

Having the displacement of every particle \{$\B d_i\}_{i=1}^N$, we consider $K$ annuli of fixed width, limited by circles of radii $\{R_k<R_{k+1}\} _{k=1}^{K-1}$ such that $R_{k+1}-R_k=\Delta$ and  $R_K=R_{\rm out}$. In each annulus we compute the averages $f_k^{(m)}$ and $g_k^{(m)}$ which are defined by: 
\begin{eqnarray}
	f_k^{(m)} \!\!&\equiv& 2\pi (k\!+\!1/2)\Delta \sum_i \B d_i \cdot \hat {\B r} \sin (m\theta_i) \ , ~ m=1,2,3\dots\label{deffg}\\
		g_k^{(m)} \!\!&\equiv& 2\pi (k\!+\!1/2)\Delta  \sum_i \B d_i \cdot \hat {\B r} \cos (m\theta_i) \ , ~ m=0,1,2,3\dots \nonumber
\end{eqnarray}   
The factor $(k+1/2)\Delta$ stands for the mid-radius associated with the $k'th$ annulus. In Fig.~\ref{iso} we show $g_k^{(0)}$ which is simply the radial component of the displacement field $d_r(r)$ multiplied by $2\pi r$. In Figs.~\ref{sin} we show
$f_k^{(m)}$ for $m=1,2$ and 3. Similarly in Figs.~\ref{cos} we show
$g_k^{(m)}$ for $m=1,2$ and 3. The data are shown as circles, and the continuous line
pertains to the theoretical curve that is explained in the next section. 
%%%%%%%%%%%%%%%%%%
 %%%%%%%%%%%%%%%%%%
 %%%%%%%%%%%%%%%%%%
%%%%%%%%%%%%%%%%%%%%%%%%%%%%%%%%%%

\section{Theory}
\label{Theory}
\subsection{Michell Solutions}
To understand the numerical results we turn now to the classical Michell solutions \cite{99Mic} for the displacement field in polar coordinates. These solutions refer precisely to the geometry that we have used above. In two dimensions, 
when no body forces are present, the stresses can be expressed through the Airy stress function $\chi$. 
Strain compatibility equations restrain the Airy stress function to satisfy the bi-harmonic equation \cite{Landau,53Mus},
%%%%%%%%%%%%%%%
\begin{equation}\label{S1}
\nabla^{2} \nabla^{2}\chi= 0. 
\end{equation}
%%%%%%%%%%%%%%%%%%
Since stresses and displacements must be single-valued and continuous, any solution of the displacements and the stress functions must be periodic functions of $\theta$. This was used by Michell to write a general solution for $\chi$ in Fourier series. Using this series one can directly compute the radial component of the displacement field in the form \cite{02Bar,99Mic,53Mus}
%%%%%%%%%%%%%%%%
\begin{align}\label{S8}
\vec{d}\cdot \hat{r} &= A_{0}r + B_{0}r(\ln(r)-1) + C_{0}r^{-1} + D_{0}\theta \nonumber \\ 
  &+ \left[ A_{1} + A_{1}^{\prime}\theta + B_{1}r^2 +  C_{1}r^{-2} + D_{1}\ln(r)\right] \sin(\theta) \nonumber \\
  &+ \left[ E_{1} + E_{1}^{\prime}\theta + F_{1}r^2 +  G_{1} r^{-2} + H_{1}\ln(r)\right]\cos(\theta) \nonumber  \\ 
  &+ \sum_{n=2}^{\infty} \left[ A_{n}r^{n-1} +\frac{B_{n}}{r^{n-1}}  + C_{n}r^{n+1}+ \frac{D_{n}}{r^{n+1}} \right]\sin(n\theta) \nonumber \\
  &+ \sum_{n=2}^{\infty} \left[E_{n}r^{n-1} +\frac{F_{n}}{r^{n-1}}  + G_{n}r^{n+1}+ \frac{H_{n}}{r^{n+1}} \right]\cos(n\theta)  \ .
\end{align}
%%%%%%%%%%%%%%%%
Needless to say, this is the most general solution; depending on the geometry and the boundary conditions, only few or more components may survive as the actual solution of our problem.

Integrating Eq.~(\ref{S8}) over the angle results in the angle-averaged radial component of the displacement field,
\begin{equation}
d_r(r)\equiv \oint_{0}^{2\pi} \vec{d}\cdot \hat{r} d\theta = A_{0}r + B_{0}r(\ln(r)-1) + C_{0}r^{-1} \ .
\label{radial}
\end{equation} 
In our set up we have only two boundary conditions, i.e. that the radial component of the displacement field equals $d_0$ at $r_{\rm in}$ and 0 at $r_{\rm out}$. Here we have however three unknowns, so to reach a unique solution we need another constraint. If we assume that the response is purely linear in the strain perturbation, we can exclude any monopole or  multipole in the solution, and this condition will result
in $B_0$ being zero in Eq.~(\ref{radial}). Applying then the two boundary conditions results in the classical solution\cite{21LMMPRS,22KMPS,22BMP}
\
\begin{equation}
d^{\rm lin}_{r}(r) = d_{0}\left( \frac{r^2 -r_{out}^2}{r_{rin}^2 -r_{out}^2}\right) \frac{r_{in}}{r}\ .
\label{classical}
\end{equation}
The numerical solution as shown above indicate that the last assumption is unwarranted, and that modes that are not allowed in a purely linear solution are readily excited. We therefore cannot cancel a-priori the $B_0$ term in Eq.~(\ref{S8}), and we will therefore consider now the angular integrals over the various contribution in that equation in full detail. In accordance with Eqs.~(\ref{deffg}) we will calculate the following integrals:
\begin{itemize}
	\item 
	The zeroth-order $(n=0)$ Fourier integral is 
%%%%%%%%%%%%%%%%
\begin{equation}\label{S10}
I_{0}(r) = \oint_{0}^{2\pi} \vec{d}\cdot \hat{r} dl = A_{0} + B_{0}r^{2} + C_{0}r^{2}(-1+\ln(r)),
\end{equation}
%%%%%%%%%%%%%%%%
where $dl = rd\theta$. 

\item Similarly, the first four integrals $(n=1,2,3,4)$ for the sine component are given by the following equations
%%%%%%%%%%%%%%%%
\begin{align}\label{S11}
I_{1}(r) &= \oint_{0}^{2\pi} \vec{d}\cdot \hat{r} \sin(\theta) dl = A_{1}r + B_{1}r^{3} + C_{1}r\ln(r) +\frac{D_{1}}{r} \nonumber \\
I_{2}(r) &= \oint_{0}^{2\pi} \vec{d}\cdot \hat{r} \sin(2\theta) dl = A_{2} + B_{2}r^{2} + C_{2}r^{4} +\frac{D_{2}}{r^{2}} \nonumber \\
I_{3}(r) &= \oint_{0}^{2\pi} \vec{d}\cdot \hat{r} \sin(3\theta) dl = A_{3}r^{3} + \frac{B_{3}}{r} + C_{3}r^{5} +\frac{D_{3}}{r^{3}} \nonumber \\
I_{4}(r) &= \oint_{0}^{2\pi} \vec{d}\cdot \hat{r} \sin(4\theta) dl = A_{4}r^{4} + \frac{B_{4}}{r^2} + C_{4}r^{6} +\frac{D_{4}}{r^{4}},
\end{align}
\end{itemize}
where the coefficients $A_{i},B_{i},C_{i},D_{i}$ are the material dependent parameters related to the stress components. The Fourier integrals for the cosine components have exactly the same form as sine components but with different coefficients, say $E_{i},F_{i},G_{i},H_{i}$. We denote the cosine integrals as $J_1$-$J_4$ in parallel to Eq.~(\ref{S11}).

\subsection{Comparison with simulations}

In our numerical calculations we observe that only a few lower-order $(n\leq 4)$ Fourier components are appreciably excited by the very small inflation of the inner boundary. Hence it is enough to compare only a few lower-order $(n\leq 4)$ Fourier integrals from the Michell's solution to the numerics as summarized by Eqs.(\ref{deffg}).

In Figures \ref{iso}, \ref{sin} and \ref{cos} we show, in continuous lines, the functional forms Eq.~(\ref{S11}) and their cosine counteraparts, with the coefficients fitted to the data. The fits are excellent, supporting our proposition that the Michell solutions provide the correct basis for the numerically computed displacement field. We also note that Michell solutions with $n>4$ are not appreciable excited, although they may very well become important for larger inflations $d_0$.

\section{Data reconstruction using the Michell solutions}
\label{recon}
\begin{figure}
	\includegraphics[width=0.35\textwidth,angle=-90]{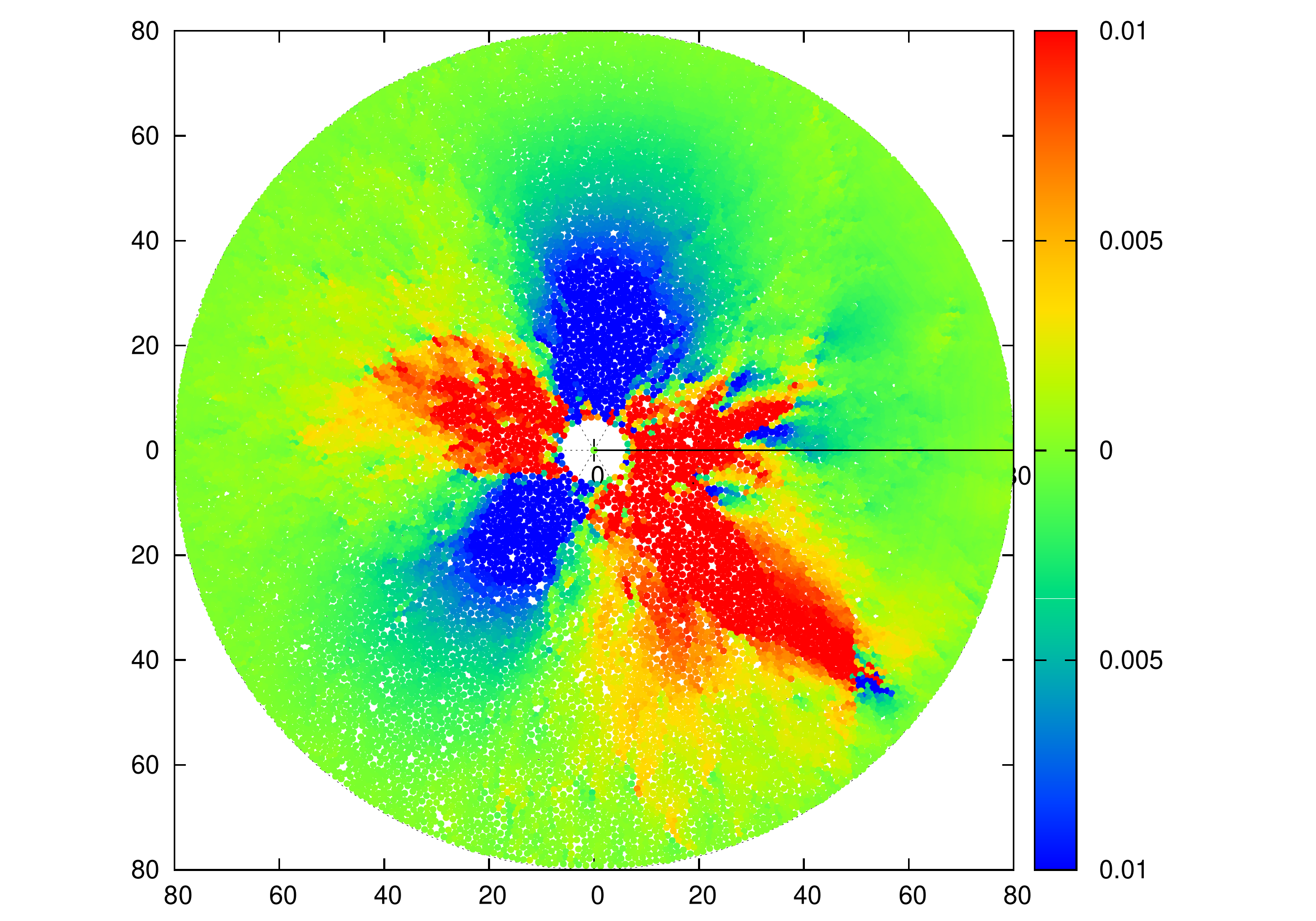}
		\includegraphics[width=0.35\textwidth,angle=-90]{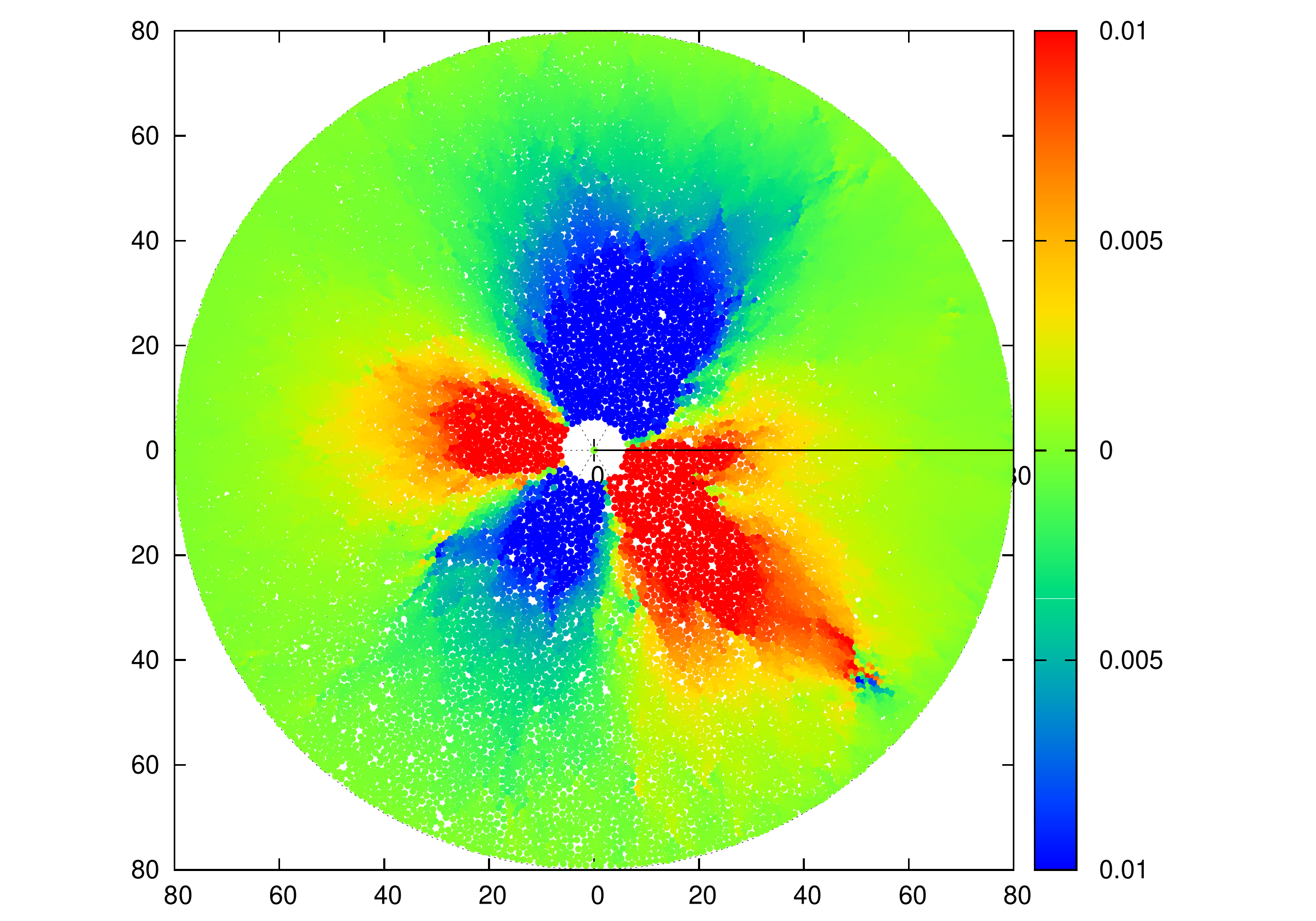}
	\caption{Upper panel: The symmetry breaking of the radial component of the displacement field, see Eq.~(\ref{S12}) for definition, with positive sign for outgoing and negative sign for incoming vector.
	Lower panel: The radial component of the response to inflation containing only Fourier modes with $n=1,2,3$ and 4, see Eq.~(\ref{Fourin})}
	\label{symbreak}
\end{figure}

Finally, to connect the symmetry-breaking modes to the Michell solutions, we will subtract in each of our $K$ annuli the angle-averaged radial component of the displacement field, Eq.~(\ref{classical}), from the displacement of each particle, and define a new, purely non-radial displacement data $D_i$: 
\begin{equation}\label{S12}
\B D_i=  \B d_i -  \hat{\B r}_{i} d_r^{\rm lin} \big((k+1/2)\Delta\big)  \ , i\in \text{k'th circular shell} \  .
\end{equation}
We plot these data such that for every particle $i$ we compute the radial component of $\B D_i$, thus determining a positive or negative sign depending on this component being outgoing from the center or incoming towards the center. Then we plot the magnitude of $\B D_i\cdot \hat{r_i}$ at every $r_i,\theta_i$ with color code that includes the sign. The resulting image associated with the data presented in Fig.~\ref{simulation} is presented in the upper panel of Fig.~\ref{symbreak}.
%%%%%%%%%%%%%%%%%%%%%%%%%%%%%%%%%%%
%%%%%%%%%%%%%%%%%%
%%%%%%%%%%%%%%%%%%%%%%%%%%%%%%%%%%
This presentation of the data underlines the importance of the symmetry-breaking components in the response to the purely radial inflation. 

Finally we demonstrate that the displacement field presented in Fig.~\ref{symbreak} can be obtained, to a very good approximation,
by inflating the inner boundary not radially, but by summing up
Fourier components. To this aim we perform now an inflation of the inner boundary according to
\begin{equation}
	r_{\rm in} \to 	r_{\rm in}+ \sum_{n=1}^4 \alpha_n \cos (n\theta) + \sum_{n=1}^4 \beta_n \sin (n\theta)  \ ,
	\label{Fourin}
\end{equation}
where
\begin{equation}
	\alpha_n\equiv \int_{r_{\rm in}}^{r_{\rm out}} J_n(r) dr \ , \quad
		\beta_n\equiv \int_{r_{\rm in}}^{r_{\rm out}} I_n(r)  dr \ .
\end{equation}
Needless to say, this form depends on the realization since the integrals $I_n$ and $J_n$ vary from system to system, reflecting the randomness in the amorphous configurations. Plotting the resulting signed magnitude of the displacement field (without any need of subtraction of a nonexistent radial part), we obtain the data shown in the lower panel of Fig.~\ref{symbreak}.
The correspondence between the upper and lower panels of Figs.~\ref{symbreak} is obvious.

%%%%%%%%%%%%%%%%%%%%%%%%%%%%%%%%%%%%%%
\section{Discussion}
\label{discuss}
To understand the reason for the mode-coupling of the radial inflation to the Michell solution we need to recall that on local scales the disorder in our amorphous solid is never ``small". In fact, it can be quantified by a typical length, say $\xi$, which is expected to diverge in granular packing when the unjamming point is approached \cite{05SLN,18Ler}.  At finite pressure this length is always finite, and we expect that small perturbations applied on larger scales should obey linear elasticity theory. In the present context this implies that by increasing $r_{\rm in}$ beyond $\xi$ the mode-coupling observed above should disappear.  
%%%%%%%%%%%%%%%%%%%%%%%%
%%%%%%%%%%%%%%%%%%
\begin{figure}
	\includegraphics[width=0.35\textwidth,angle=-90]{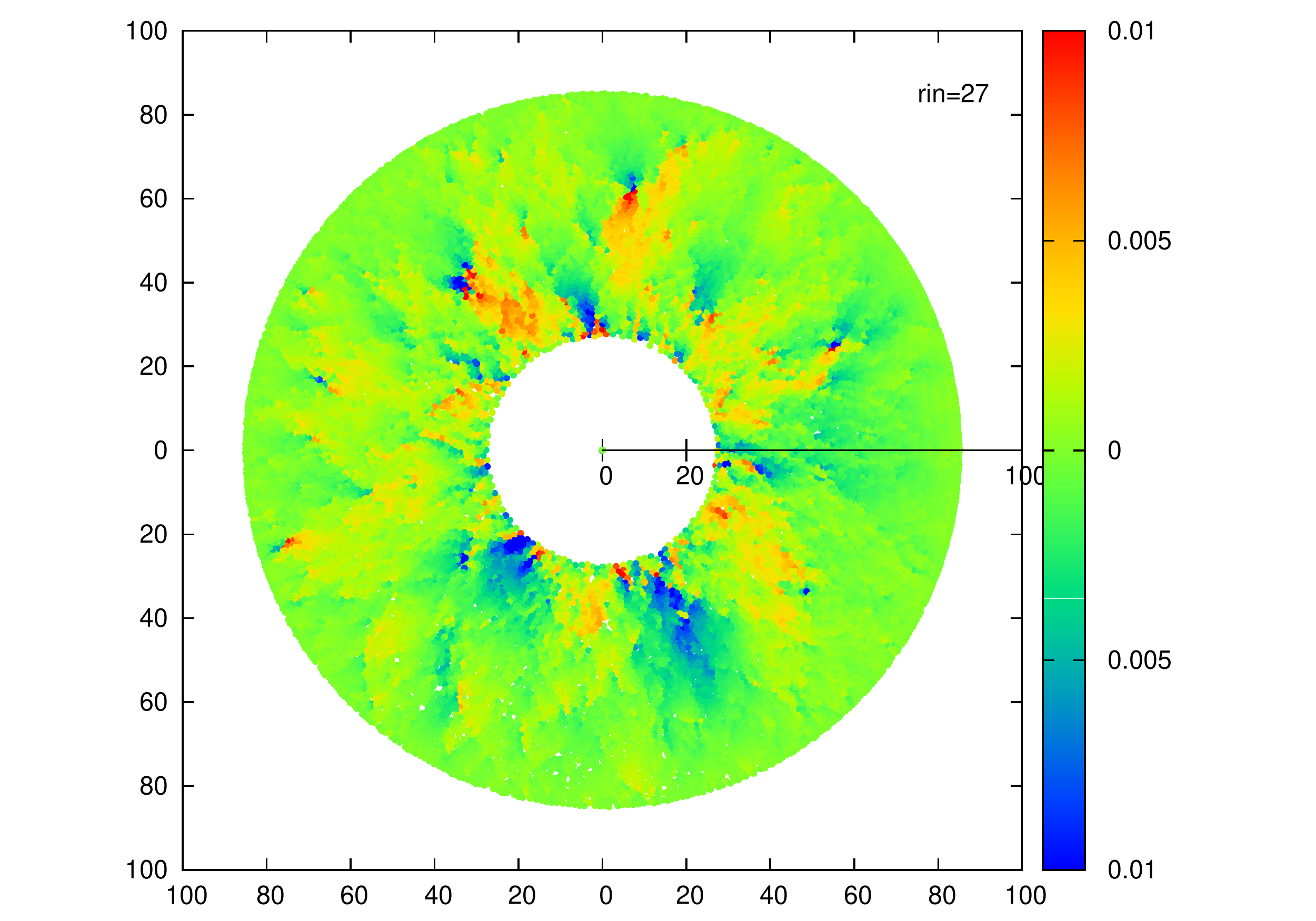}
	\caption{Magnitude of the displacement field resulting form a minute
			purely radial inflation $r_{\rm in} \to r_{\rm in}+\delta$ where $r_{\rm in}=27$,
			$r_{\rm out}=84$ and $\delta=10^{-6}$. The non-radial response is reduced now to almost random noise. }
	\label{largerin}
\end{figure}
%%%%%%%%%%%%%%%%%%%%%%%%%%%%%%%%%%
This is demonstrated in Fig.~\ref{largerin}. Taking now $r_{\rm in}=27$ the mode-coupling effect becomes suppressed. Computing the integrals shown above in Figs.~\ref{sin} and \ref{cos} results in random numbers that cannot be fitted to the expected Michell solutions at all. 

\begin{figure}
	\includegraphics[width=0.35\textwidth,angle=-90]{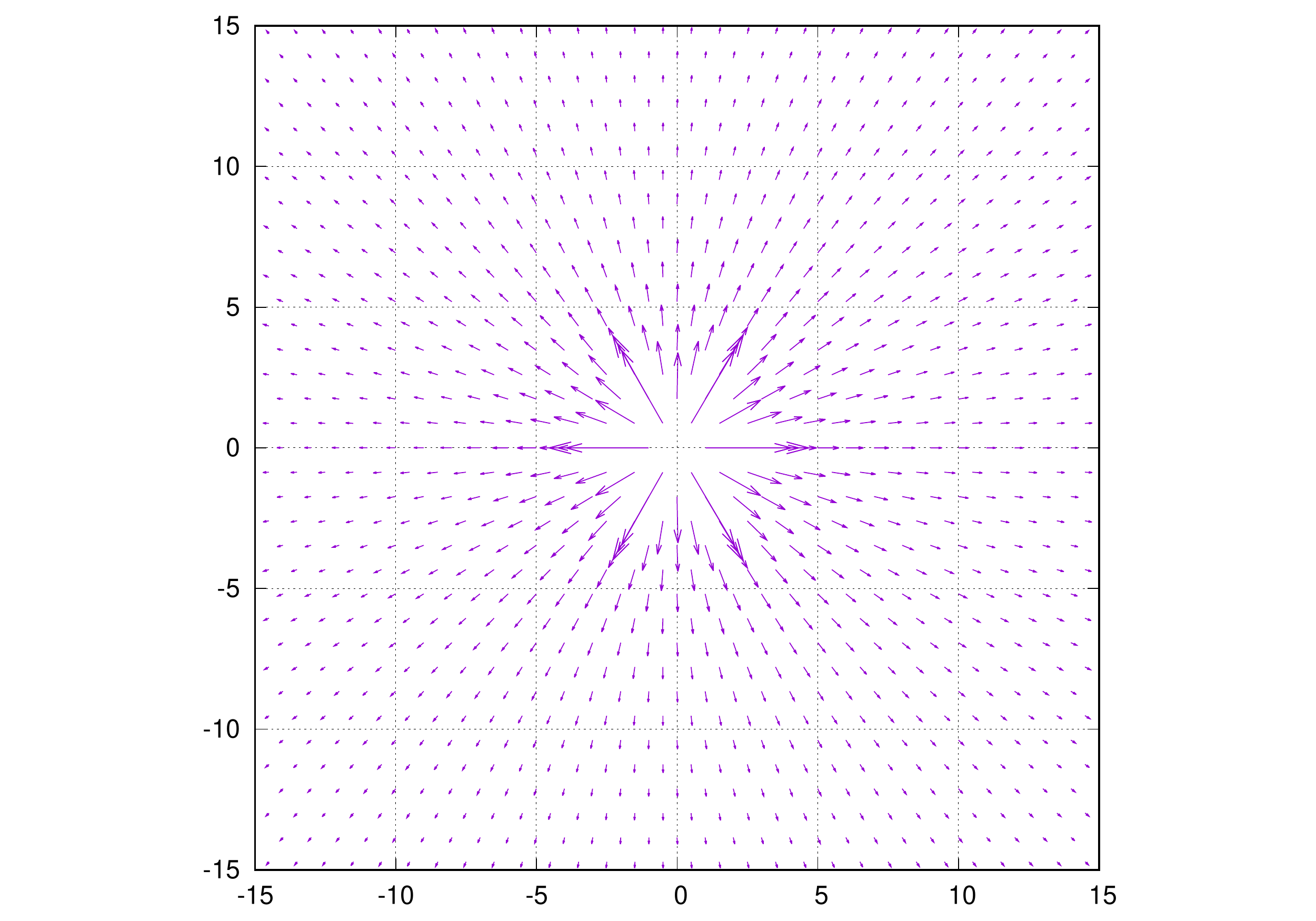}
	\includegraphics[width=0.35\textwidth,angle=-90]{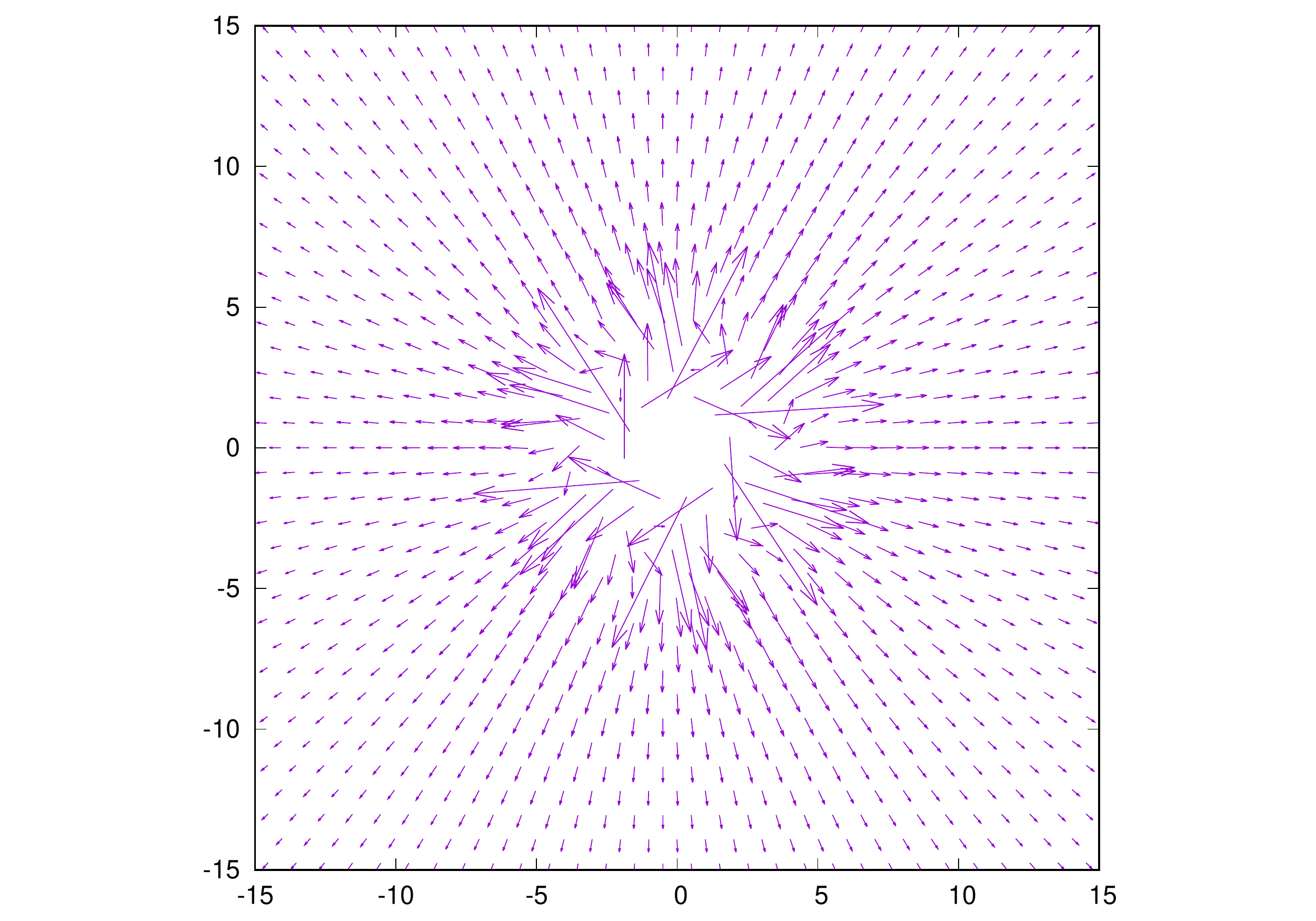}
	\caption{Upper panel: the displacement field resulting form a minute
		purely hexagonal inflation into a crystalline hexagonal configuration. The symmetry is perfectly retained. Lower panel: the displacement field induced by a minute inflation of a radial inner boundary into  a crystalline hexagonal configuration. The  response is now disordered only close to the inner boundary and symmetry is restored immediately after, without exciting nonlinear Michell solutions.}
	\label{hexagon}
\end{figure}

To complete the identification of the disorder as an inducer of symmetry breaking, we repeated the simulations of inflating an inner boundary into a perfectly ordered hexagonal crystal. When we choose the inner boundary to be a hexagon, we obtain the displacement field exhibited in the upper panel of Fig.~\ref{hexagon}.  
%%%%%%%%%%%%%%%%%%
%%%%%%%%%%%%%%%%%%%%%%%%%%%%%%%%%%
The symmetry is perfectly retained in this case. On the other hand, if we choose the inner boundary to be circular, we obtain the displacement field shown in the lower panel of Fig.~\ref{hexagon}. Here the clash between the circular and hexagonal symmetry results in a disordered displacement near the inner boundary, but this disorder is quickly suppressed in favor of an ordered displacement field in the farther field. No nonlinear mode-coupling to Michell solutions is observed.

In summary, we have shown that the existence of randomness in amorphous solids can lead to appreciable mode-coupling effects when the scale of the applied strain is within the disorder length. Then modes that appear in the Michell analysis are excited even by perturbation of different symmetry. Radial perturbation can be mode-coupled to higher order Fourier modes. Linear elasticity theory cannot be trusted and more general solutions for the displacement field should be considered. This is before plasticity effects are taken into account. These are expected to show up when the amplitude of the perturbation increases, leading to other interesting and non-trivial breakdowns of elasticity theory \cite{21LMMPRS,22KMPS,22BMP,22MMPRSZ}. 

The phenomenon discussed in this paper appears relevant for modeling  plasticity in amorphous solids as well. As is well known, plastic events are generically Eshelby quadrupoles \cite{54Esh} carrying their own eigen-strain with core sizes that are typically smaller than the disorder length. Assuming therefore that their influence on neighboring region of amorphous matter can be modeled by the {\em linear} Eshelby kernel may need careful reconsideration. 

We thank Michael Moshe for various useful discussions on the research presented here. This work had been supported in part by ISF under grant \#3492/21 (collaboration with China) and the Minerva Center for ``Aging, from physical materials to human tissues" at the Weizmann Institute.

\bibliographystyle{unsrt}
\bibliography{MetaMat,Phasetran}
\end{document}